\documentclass[12pt]{article}
\usepackage{amsmath,latexsym,amssymb,epsfig,psfrag}  
\usepackage{axodraw} 
\usepackage{comment}

\def\ltsim{\lower3pt\hbox{$\, \buildrel < \over \sim \, $}}  
\def\gtsim{\lower3pt\hbox{$\, \buildrel > \over \sim \, $}}  
\newcommand{\be}{\begin{equation}}  
\def\ga{\mathrel{\raise.3ex\hbox{$>$\kern-.75em\lower1ex\hbox{$\sim$}}}}  
\def\la{\mathrel{\raise.3ex\hbox{$<$\kern-.75em\lower1ex\hbox{$\sim$}}}}

\newcommand{\de}{\partial}  
\makeatletter   
\@addtoreset{equation}{section}   
\makeatother

\def\simgt{\stackrel{>}{{}_\sim}}  
\begin{document}  
  
\baselineskip=16pt   
\begin{titlepage}  
\begin{center}  
\hfill{{\bf UAB-FT-547}}\\
\hfill{{\bf SPhT-T03/067}}\\
\hfill{{\bf DFPD-03/TH/21}}

\vspace{0.5cm}  
  
\large {\sc \Large Supersymmetry Breaking with Quasi-localized \\
\Large  Fields in Orbifold Field Theories}

\vspace*{5mm}  
\normalsize  
  
{\bf G.~v.~Gersdorff~\footnote{gero@ifae.es}, 
L.~Pilo~\footnote{pilo@shpht.saclay.cea.fr}, 
M.~Quir\'os~\footnote{quiros@ifae.es}, 
D.~A.~J. Rayner~\footnote{rayner@pd.infn.it}, 
A.~Riotto~\footnote{antonio.riotto@pd.infn.it}}   

\smallskip   
\medskip   
\it{~$^{1,\,3}$~Theoretical Physics Group, IFAE}\\ 
\it{E-08193 Bellaterra (Barcelona), Spain}

\smallskip   
\medskip
\it{~$^2$~Service de Physique Th\'eorique, CEA/DSM/PhT\\
Unit\'e de recherche associ\'ee au CNRS, CEA/Saclay\\
91191 Gif-sur-Yvette c\'edex, France} 

\smallskip   
\medskip
\it{$^3$~Instituci\'o Catalana de Recerca i Estudis Avan\c{c}ats (ICREA)}

\smallskip    
\medskip  
\it{~$^{4,\,5}$~Department of Physics and INFN,\\
Sezione di Padova, via Marzolo 8,
I-35131 Padova, Italy}


\vskip0.6in \end{center}  
   
\centerline{\large\bf Abstract}  

\noindent
We study the Scherk-Schwarz supersymmetry breaking in five-dimensional
orbifold theories with five-dimensional fields which are not strictly
localized on the boundaries (quasi-localized fields).  We show that
the Scherk-Schwarz (SS) mechanism, besides the SS parameter $\omega$,  
depends upon new parameters,
e.g. supersymmetric five-dimensional odd mass terms, governing the
level of localization on the boundaries of the five-dimensional fields
and study in detail such a dependence. Taking into account radiative 
corrections, the value of $\omega$ is dynamically allowed to acquire any
value in the range $0< \omega < 1/2$.



\vspace*{2mm}   
  
\end{titlepage}  
  
\section{\sc Introduction}  \label{sec:intro}

Gauge theories in more than four dimensions are interesting due to the
appearance of new degrees of freedom whose dynamics can spontaneously
break the symmetries of the theory. In particular, the dynamics of
Wilson lines, which become physical degrees of freedom on a
multiply-connected manifold and parametrize degenerate vacua at the
tree level, can lift the vacuum degeneracy after quantum corrections
are included. This is the so-called Hosotani
mechanism~\cite{hosotani}.  On the other hand, on multiply-connected
manifolds, non-trivial boundary conditions imposed on fields can
affect the symmetries of the theory. This mechanism was proposed long
ago by Scherk and Schwarz (SS) for supersymmetry breaking~\cite{ss},
which remains one of the open problems of the theories aiming to solve
the hierarchy problem by means of supersymmetry.  In five-dimensional
(5D) theories compactified on the orbifold $S^1/\mathbb Z_2$, the
softness of the SS supersymmetry breaking was demonstrated by explicit
calculations~\cite{explicit} and interpreted as a spontaneous symmetry
breaking through a Wilson line in the supergravity completion of the
theory~\cite{geromariano,Quiros:2003gg}. This means that the Hosotani
mechanism to break local supersymmetry and the SS mechanism are
equivalent~\cite{anomalies}.  In particular such mechanisms to break
supersymmetry arise from the vacuum expectation value (VEV) of an
auxiliary field $\langle V_5^1+i V_5^2\rangle$ of the 5D off-shell
supergravity multiplet that appears in the low-energy effective theory
as the auxiliary field of the $N=1$ radion supermultiplet
\begin{equation}
 \mathcal R=\left(h_{55}+i B_5,\psi^2_{5L},V_5^1+i V_5^2\right)\, ,
  \label{radion}
\end{equation}
where $h_{MN}$ is the 5D metric, $B_M$ the graviphoton, and $\psi^i_M$
the gravitino, where the indices $i=1,2$ transform as a doublet of the
$SU(2)_R$ symmetry. Making use of $SU(2)_R$ we can orientate the VEV
along, e.g.~$V_5^2$, and define the VEV
\begin{equation}\label{omega}
 \langle V_5^2\rangle=\frac{\omega}{R}
\end{equation}
in terms of a parameter $\omega$ (where $R$ is the radius of
$S^1$). The tree-level potential in the background of (\ref{omega}) is
flat, reflecting the no-scale structure of the SS breaking.  However
this degeneracy is spoiled by radiative effects. In particular for a
system of $N_V$ vector multiplets and $N_h$ hyperscalars in the bulk,
the one-loop effective potential was computed in
Ref.~\cite{geromarianotoni} to be
\begin{equation}
\label{pot0}
V_{eff}(\omega)=\frac{3(2+N_V-N_h)}{64\pi^6 R^4}\left[\text{Li}_5
(e^{2 i\pi\omega})+\text{h.c.}\right]\, ,
\end{equation}
where the polylogarithm function is defined as
\begin{equation}
\label{poly}
\text{Li}_n(x)=\sum_{k=1}^\infty\frac{x^k}{k^n}\ .
\end{equation}
Notice that potential (\ref{pot0}) has a minimum at $\omega=0$
($\omega=1/2$) for $N_h>2+N_V$ ($N_h<2+N_V$) depending on the
propagating bulk matter, while it does not depend on the $N=1$
supersymmetric matter localized at the orbifold fixed-points $y=0, \pi
R$.

The localization properties of KK wave functions can be altered by adding a 
bulk mass term (possibly with a non-trivial profile in the fifth dimension) 
to achieve (quasi)-localization of bulk fields at the
fixed-point branes~\cite{georgi}. In particular, we will be interested 
in 5D hypermultiplets with odd-parity bulk masses, where such mass terms 
can also be thought as localized  
Fayet-Iliopoulos (FI) terms corresponding to a $U(1)$ gauge group under which 
hypermultiplets are 
charged. These FI terms, even when absent at tree-level, are generated 
radiatively~\cite{nilles}. 
This issue was analyzed in detail in~\cite{bar} where it was
shown that the 5D supergravity extension of a FI term could be made
for a flat theory where the gravitino has zero $U(1)$ charge,
i.e.~where the R-symmetry is not gauged. Moreover an odd mass term can
exist even in the absence of a FI term for global supersymmetry. In
the supergravity extension it should follow from the graviphoton $B_M$
gauging, the mass of each hypermultiplet being proportional to its
gravicharge $Q_B$. So, in the absence of a FI term (in which case the
gravitino is coupled to the graviphoton but not to the $U(1)$ gauge
boson) or even if there is no $U(1)$ factor, an odd supersymmetric
mass can be introduced for gravicharged hypermultiplets. This provides
a very general mechanism for localization of bulk hypermultiplets.

In this letter we will study the Hosotani mechanism in 5D theories
compactified on the orbifold $S^1/\mathbb Z_2$ in the presence of
quasi-localized fields which are not strictly localized at the
boundaries.  Note that fields which are strictly localized to the
boundary fixed points with delta functions are four-dimensional fields
and therefore do not couple directly to the Wilson line: as such, they
cannot affect the dynamics of the Wilson line. However, if
five-dimensional fields are localized on the boundaries by some
mechanism, e.g. by a five-dimensional mass term, they can still have
an influence on the dynamics of the Wilson line. In particular, we
expect that the selection (at the quantum level) of the vacuum of the
underlying gauge theory depends on some new parameter(s) quantifying
the level of localization on the boundaries of the five-dimensional
fields.  If this parameter is a five-dimensional mass term $M$ and if,
for $\left|M\right|\rightarrow \infty$, strict localization is
attained, we expect the effect on the Wilson line dynamics to
disappear in the limit of very large $\left|M\right|$.  Here we will
restrict ourselves to study the effects of quasi-localized fields on
the SS-mechanism for supersymmetry breaking. In general, we expect the
SS-supersymmetry breaking parameter to depend upon the new parameter
$\left|M\right|$.  We will leave the analysis of such effects on the
spontaneous symmetry breaking in five-dimensional gauge theories for a
future publication~\cite{appear}.

This letter is organized as follows. In section 2 we will give a short
review of the SS and Hosotani mechanisms in a 5D orbifold. In section
3 we will calculate the Kaluza-Klein (KK) mass spectrum and
corresponding wave functions for hypermultiplets with arbitrary
SS-supersymmetry breaking parameter $\omega$ and odd supersymmetric
bulk masses $M$. Section 4 is devoted to the actual computation of the
effective potential and the dynamical determination of the value of
the VEV of the SS-parameter is done in section 5.  Finally in section
6 we draw our conclusions.

\section{\sc Scherk-Schwarz/Hosotani breaking on an orbifold}  
 \label{sec:ss}

In this section we will review and compare the Scherk-Schwarz and
Hosotani symmetry breaking mechanisms in 5D orbifold models.  We will
consider the spacetime manifold ${\cal M} = \mathbb{R}^4 \times {\cal
C }$, where the compact component ${\cal C}$ is a coset (singular)
space $\mathbb{R}/\mathcal G$.  In our case ${\cal G}$ is the
semi-direct product ${\cal G} = \mathbb{Z} \ltimes
\mathbb{Z}_2$. Calling $\tau$ and $\zeta$ the generators of
$\mathbb{Z}$ and $\mathbb{Z}_2$ respectively, they act as
\begin{equation}
 \tau(y) = y + 2 \pi R \, , \qquad   \zeta(y) = -y \quad .
\end{equation}
The action of $\mathcal G$ has two fixed points $0$ and $\pi R$ and
the resulting space is an orbifold.  A generic field $\phi$ is defined
on ${\cal M}$ by modding out the action of $\mathcal G$,
\begin{gather}
 \phi(x,\tau(y) \, ) = T \, \phi(x,y) \quad ; \label{twist} \\
  \phi(x,\zeta(y) \, ) = Z \, \phi(x,y) \quad ; \label{refl}
\end{gather}
where $T$ and $Z$ are global (local) symmetry transformations
represented by suitable matrices acting in the field space. From $\tau
\cdot \zeta \cdot \tau(y) = \zeta(y)$, we obtain the following consistency
relation $T Z T = Z$. The element $\zeta^\prime = \zeta \tau$ of
${\cal G}$ generates a second $\mathbb{Z}_2$ transformation and ${\cal
G}$ can be equivalently considered as generated by $\zeta$ and
$\zeta^\prime$. In general the action of $\zeta^\prime$ in field space
$Z^\prime = Z \, T$ does not commute with $Z$.  The modding-out in
Eqs.~(\ref{twist})-(\ref{refl}) can be used to break softly some (or
all) of the symmetries involved in the non-trivial boundary
conditions.  

As for matters fields, ${\cal G}$ acts also on the gauge fields
$V_{{}_M}$, and if $g$ is a generic element of ${\cal G}$, we have
\be 
V_{{}_M}^a(x, g(y)) =
c_{{}_M} \, {\Lambda_g}^{ab} \, V_{{}_M}^b(x, y) \quad (\text{no sum
on }M) \, , \qquad c_{{}_M} = \begin{cases} 1 & M = \mu \\ \de_y g(y)
& M = 5 \end{cases} .
\label{gft}
\end{equation}   
Requiring that the covariant derivative of matter fields transforms
consistently, e.g.
\begin{equation}
\begin{split} 
\phi(x, g(y)) &= M_g \, \phi(x,y) \, \quad \\
\Rightarrow \, D_{{}_M} \phi(x,g(y)) &= c_{{}_M} \, M_g \, D_{{}_M}
\phi(x,y) \quad (\text{no sum on }M) ;
\end{split}
\end{equation}
we get
\be \label{automorphism}
M_g \, T_a \, M_g^{-1} = T^\prime_a = {\Lambda_g}^{ab} \, T_b \qquad .
\end{equation}
Thus, as a result, ${\cal G}$ acts on the Lie algebra of the gauge
group $G$ as an automorphism.  Finally, imposing that a gauge
transformation does not alter (\ref{gft}) it follows that the
associated gauge parameters $\xi^a$ satisfy the relation
\be 
\xi^a(x, g(y)) = {\Lambda_g}^{ab} \,
\xi^b(x,y) \quad .
\end{equation} 

Let us now focus on the case $G=SU(2)$ and matter fields $\phi$
transforming as doublets.  One can represent the $T$ and $Z$ symmetry
transformations in field space as
\begin{equation}
 T = e^{2 \pi i \, \omega \, \sigma_2} \, , \qquad Z = Z_{\text{Lor}}
  \otimes \sigma_3 \, , \qquad  Z_{\text{Lor}}= 
   \begin{cases}
    1 & \text{5D Scalars} \\
    i \gamma^5 & \text{5D Dirac Fermions} 
   \end{cases} 
    \label{eq:stwist} \quad ;
\end{equation}
where the matrix $Z_{\text{Lor}}$ acts on the Lorentz indexes.  A
non-trivial twist $T$ triggers Scherk-Schwarz breaking.  However, when
the $SU(2)$ symmetry is gauged, the SS-mechanism is equivalent to
spontaneous breaking by the Hosotani mechanism.  Making the choice
(\ref{eq:stwist}) for the matter fields, we have that the automorphism
in Eq.~(\ref{automorphism}) is given by
\be
\Lambda_\zeta = \begin{pmatrix}
-1 & 0 & 0& \\ 0 & -1 & 0 \\ 0 & 0 & 1 \end{pmatrix} \,  \quad .  
\end{equation}   
Notice that when $\omega \neq 0$, $SU(2)$ is completely broken, while
the case $\omega=0$ corresponds to the breaking pattern $SU(2)
\rightarrow U(1)$. The case $\omega = 1/2$ is special: $[Z, Z^\prime]
= 0$, $Z= \sigma_3$, $Z^\prime = \epsilon \sigma_3, \quad \epsilon^2
=1$, and the KK modes can be classified according to the two
independent parities\footnote{This case is often considered in the
literature without referring to SS breaking.}.
 
In the Hosotani basis $\varphi$ the gauge potential has a non trivial
VEV $\langle V_{{}_M} \rangle = - \delta^5_{{}_M} \, \omega R^{-1} \,
\sigma_2$ and the twist $T$ is trivial.  The SS basis $\phi$ and the
Hosotani basis $\varphi$ are related by the following {\it
non-periodic} gauge transformation
\be 
\phi = e^{i
\omega  y \sigma_2/R} \, \varphi \quad .
\end{equation}
In the SS basis the field $\phi$ satisfies twisted boundary conditions 
and the background gauge field is vanishing.  Also note that a constant 
VEV for $V_5^2(x,y)$ is allowed only if $V_5^2(x,y)$ is even, and only 
the part of the breaking corresponding to the twist can be viewed as 
spontaneous.

\section{\sc Kaluza-Klein mass spectrum}  \label{sec:KK}

Let us now consider a supersymmetric hypermultiplet $(\varphi^i,\psi)$
in five-dimensions with a localizing odd-parity mass term $M(y)$. Its
Lagrangian is
\begin{equation}
\label{lag}
\mathcal L=\left|\mathcal D_M\varphi\right|^2+i\bar\psi \gamma^M
D_M\psi+ M(y)\bar\psi\psi-M^2(y)\left|\varphi\right|^2+\partial_5M(y)
\left(\varphi^\dagger \sigma_3 \varphi\right) \quad .
\end{equation}
where $M(y)=\eta(y)M$, and $\eta(y)$ is the sign function on $S^1$
with period $\pi R$, which is responsible for the localization of the
supersymmetric hypermultiplet.  Working in the Hosotani basis, the
``covariant derivative'' is given by $\mathcal D_M=D_M+i\sigma_2
R^{-1} \omega \, \delta_{M5}$, where $D_M$ is the normal covariant derivative
with respect to the gauge group, and $\varphi=(\varphi^1,\varphi^2)^T$
is a doublet upon which the $\sigma_i$ matrices are acting.

\begin{figure}[htb]
\begin{center}
\psfrag{m01}{\hspace*{-6mm}$(m_0 R)^2$ }
\psfrag{m02}{\hspace*{-5mm}$(m_0 R)^2$ }
\psfrag{m03}{\hspace*{-6mm}$(m_0 R)^2$ }
\psfrag{m04}{\hspace*{-5mm}$(m_0 R)^2$ }
\psfrag{M1}{\hspace*{5mm}$MR$}
\psfrag{M2}{\hspace*{5mm}$MR$}
\psfrag{M3}{\hspace*{5mm}$MR$}
\psfrag{M4}{\hspace*{5mm}$MR$}
\scalebox{0.7}{ {\mbox{\epsfig{file=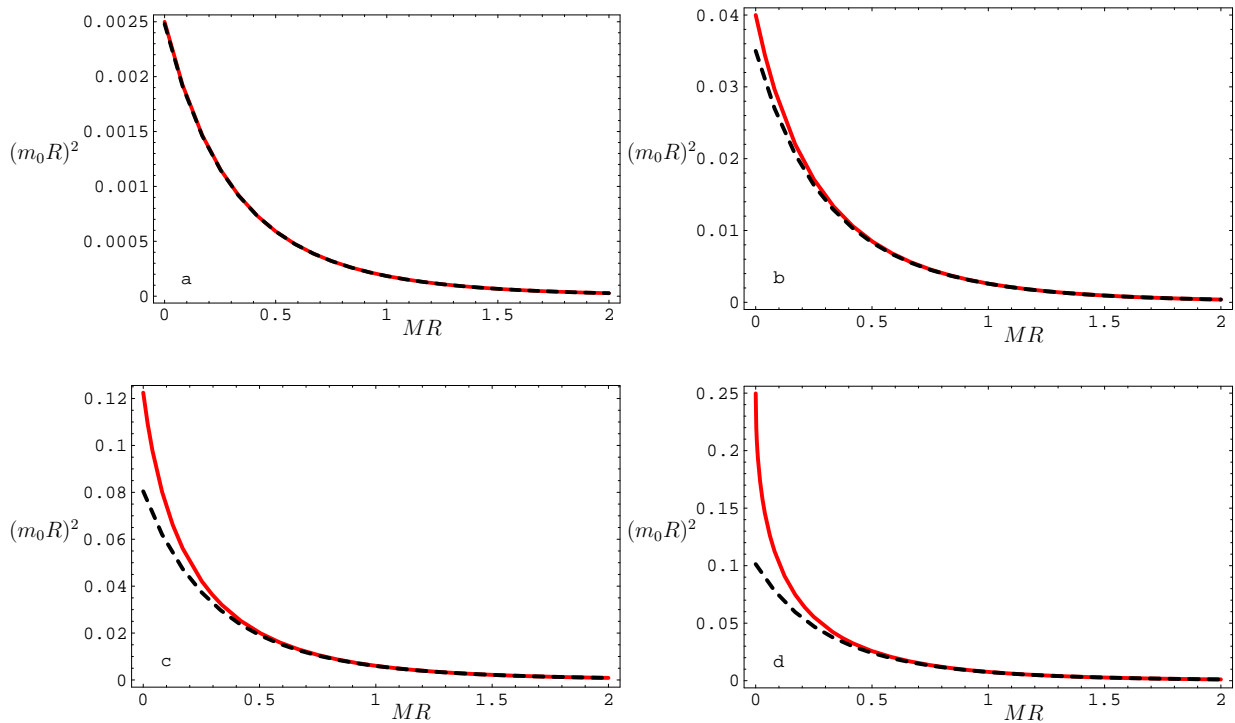}}}}
\caption{{\small  
The ``lightest'' state mass $m_0$ as a function of
$MR$ for different values of $\omega$: a) 0.05; b) 0.2; c) 0.35 and d)
0.5. Full line (red) is the numerical result from (\ref{masas}) and
dashed-line (black) the analytical approximation in (\ref{m0}).}}
\label{fig1}
\end{center}
\end{figure}

Setting $\varphi(x,y) \rightarrow e^{-i p x} \varphi(y)$, the free part of the 
equations of motion (EOM) become
\begin{equation}
\label{eomvarphi}
 \left\{\de_5^2  + m^2-M^2 \, - \frac{\omega^2}{R^{2}} \, -2 M \, 
  [\delta(y)-\delta(y-\pi
 R)]\sigma_3+ 2 i \frac{\omega}{R} \, 
   \sigma_2 \, \de_5 \right\}\varphi=0 \quad ;
\end{equation}
where   we   have   used   the   on-shell   condition   $p^2 = m^2$. 
Integrating over  a small interval around $y=0,\pi
R$ we obtain the following  boundary conditions for the even component
$\varphi_1$:
\begin{equation}
 \begin{split}
   \partial_5\varphi_1(0^+) =& M\varphi_1(0) \quad ; \\
   \partial_5\varphi_1(\pi^-)=& M\varphi_1(\pi) \quad .
 \end{split}
  \label{bcond1}
\end{equation}
while those for the odd component $\varphi_2$ are
\begin{equation}
 \varphi_2(0)=\varphi_2(\pi)=0. 
  \label{bcodd}
\end{equation}
The solutions of (\ref{eomvarphi}) in the interval $(0,\pi R)$ subject
to the previous boundary conditions are given by
\begin{equation}
\label{relacion}
\varphi(y)=e^{-i\sigma_2\omega y/R}\phi(y) \quad ;
\end{equation}
where $\phi(y)$ is the SS-wave function given by
\begin{align}
\label{SSfunc}
\phi_1=& c \, \left(\cos\Omega y+\frac{M}{\Omega}\sin\Omega y\right) \quad ;
\nonumber\\ \phi_2=& - c \, \tan\omega\pi\left(\cot\Omega\pi
R+\frac{M}{\Omega}\right)\sin\Omega y \quad ;
\end{align}
where $c$ is a normalization constant and 
$\Omega^2=m^2-M^2$. The 4D mass spectrum is given by 
\begin{equation}
\sin^2\omega\pi=\frac{M^2+\Omega^2}{\Omega^2}\sin^2\Omega\pi R  \quad .
\label{masas}
\end{equation}
with solutions providing the mass-eigenvalues and mass-eigenfunctions
that are given in (\ref{relacion}) and (\ref{SSfunc}). Even though we
cannot solve analytically (\ref{masas}) to find the mass-eigenvalues
$m$, we can consider two interesting limits. For $\Omega^2 \gg M^2$ we have
the approximate solutions $\Omega_n\simeq (n+\omega)/R$. Most
interestingly, for $m\ll |M|$ we get a very light state with mass eigenvalue
\begin{equation}
\label{m0}
 m_0^2 \simeq M^2 \frac{\sin^{2}(\pi\omega)}{\sinh^{2}(M\pi R)} \quad .
\end{equation}
The exact numerical solutions of Eq.~(\ref{masas}) and the
approximation from Eq.~(\ref{m0}) are compared in Fig.~\ref{fig1}.

\noindent The wave function corresponding to the eigenvalue
(\ref{m0}) is given, for $M>0$, by
\begin{align}\label{wfunc+}
 \varphi_1^0=&\sqrt{2|M|R}\ \cos\left[\omega\left(\frac{y}{R}-\pi\right)
  \right] e^{M(y-\pi R)} \quad ;\nonumber\\ 
 \varphi_2^0=&\sqrt{2|M|R}\ \sin\left[\omega\left(\frac{y}{R}-\pi\right)
  \right] e^{M(y-\pi R)} \quad .
\end{align}
and for $M<0$, by
\begin{align}\label{wfunc-}
\varphi_1^0=&\sqrt{2|M|R}\ \cos\left[\frac{\omega y}{R}\right] e^{-|M| y} 
 \quad ;\nonumber\\
\varphi_2^0=&\sqrt{2|M|R}\ \sin\left[\frac{\omega y}{R}\right] e^{-|M| y} 
 \quad .
\end{align}
where we have taken the approximation $|M|\pi R>1$ that is well
justified from Fig.~\ref{fig1}. The even-parity $\varphi^{0}_{1}$
state described by the wave function (\ref{wfunc+}) is quasi-localized
at the brane $y=\pi R$, while the one described by (\ref{wfunc-}) is
quasi-localized at the brane $y=0$. Notice that these become strictly
localized in the limit $\left|M\right|\to\infty$.

The EOM of fermions $\psi$ are easily obtained from the Lagrangian
(\ref{lag}). Decomposing $\psi$ into 4D chiralities, $\psi_{L,R}$ and
assigning an even (odd) parity to $\psi_L$ ($\psi_R$) one can
decompose the fields in plane-waves, $\psi_{L,R}(x, y) =
\mathrm{exp}(-i p \cdot x) \, \varphi_{1,2\,}(y)\, \xi$, where $p^2 =
m^2$ with $m$ the mass eigenvalues, $\varphi_{1,2\,}(y)$ the mass
eigenfunctions, and $\xi$ is a constant two-component spinor. From
here on one could perform a similar analysis to the bosonic case,
taking into account that fermions are not affected by the SS-breaking,
as shown in (\ref{lag}). However this is not necessary since we can
use supersymmetry to write the final result. In fact the mass
eigenvalues are given by (\ref{masas}) for $\omega=0$, i.e.
\begin{equation}
\label{masasf}
m_n^2=\frac{n^2}{R^2}+M^2\left(1-\delta_{n0}\right) \quad .
\end{equation}
In particular the lightest eigenstate is massless, $m_0=0$. The
corresponding eigenfunctions $\varphi_{1,2}^0(y)$ can be read off from
Eqs.~(\ref{wfunc+}) and (\ref{wfunc-}) with $\omega=0$. The even
fermions $\psi_L^0$ are then quasi-localized at the branes $y=0,\pi R$
depending on the sign of the bulk mass $M$.

\section{\sc Effective potential}  \label{sec:effective}

The first step in the dynamical determination of $\omega$ is to
compute the contribution of hypermultiplets to the effective potential
$V_{eff}$.  We have
\begin{equation}
\begin{split} 
&V_{eff} = \frac{2 N_H}{2} \sum_n \int\!\! \frac{d^4p}{(2 \pi)^4} \,
  \text{ln} \left(p^2 + m^2_n \right) 
   = 2 N_H \int\!\! \frac{d^4p}{(2 \pi)^4} \, W(p) \quad ;
    \label{beff} \\
& W(p) = \frac{1}{2} \sum_n \text{ln} ( p^2 + m_{n}^{2} ) \quad ;
\end{split}
\end{equation}
where $p$ is the 4d Euclidean momentum and $N_H$ is the number of
hypermultiplets with a common odd bulk mass $M$.  Although the mass
relation (\ref{masas}) cannot be solved analytically, following the
techniques of Refs.~\cite{quiros,rothstein}, we can perform the
infinite sum to find $W(p)$ - or rather its derivative $\partial_p
W(p)$ - without requiring explicit analytical expressions for the KK
masses.  First, we convert the sum over KK mass-eigenvalues into a
contour integral $C$ around the infinite set of solutions of
Eq.~(\ref{masas}) along the real axis:
\begin{align}
  \frac{\partial \, W}{\partial \, p} =&\, p \, \sum_n \frac{1}{p^2 +
   m_{n}^{2}} =\, p \, \int_C \frac{dz}{2 \pi i} \,
   \frac{1}{\left(p^2+z^2 \right)} \, \frac{d}{dz} \text{ln}\, K(z)
   \quad ;
  \label{eq:contour} \\
 {\rm where} \hspace*{8mm} K(z) =&  \left( z^2 -M^2 \right) \, 
  \sin^2(\omega \pi) - m^2 \, \sin^2 \left( \pi R 
   \sqrt{ \vrule height 10pt width 0pt z^2 -M^2} \right) \quad . 
  \label{eq:contourk}
\end{align}
The contour $C$ encircles all the eigenvalues on the real
axis, but avoids the poles of the integrand in Eq.~(\ref{eq:contour})
at $z = \pm i \, p$.  We can deform $C$ into another contour $C'$
around the imaginary axis, with a small circular deformation close to
the poles at $z = \pm i \, p$.  Since the integrand is odd in $p$
along the imaginary axis, we find that only the residues at $z = \pm i
p$ contribute.  The final result is
\begin{equation}
 W(p, \omega) = \frac{1}{2} \, \text{ln} \left[\left(p^2+M^2 \right) 
  \cos(2 \omega \pi) - M^2 -p^2 \, 
   \text{cosh} \left(2 \pi R  \sqrt{p^2+M^2} \right) \right] + {\rm const}
 \, .
     \label{eq:w}
\end{equation} 
Note that due to 5D supersymmetry and Lorentz invariance, we cannot
write a local operator using only $V_5^2$, which implies that (at one
loop) $V_{eff}$ must be finite. Indeed the divergent part is
$\omega$-independent, which can be shown by subtracting the fermionic
part $W(p, \omega=0)$:
\begin{align}
  V_{eff} =&\, 2 N_H \, \int\!\! \frac{d^4p}{(2 \pi)^4} \, 
   \left[ \vrule height 14pt width 0pt
    W(p, \omega) - W(p, \omega=0) \right]  \label{eq:veff0} \\ 
  =&\, \frac{2 N_H}{2} \, \int\!\! \frac{d^4p}{(2 \pi)^4} \, 
   \text{ln} \left[ 1 + 
    \frac{ \left( p^2 + M^{2} \right) \,
     \sin^{2}( \omega \pi )}{ p^2 \, 
      \text{sinh}^2 \left( \pi R \, \sqrt{p^2+M^{2}} \right) } \right]
       \quad .    \nonumber
\end{align}
In the limit of vanishing bulk mass, Eq.~(\ref{eq:veff0})
recovers the standard expression for the effective potential involving
polylogarithms in Eq.~(\ref{pot0}).

It is also convenient to find an analytical expression for $V_{eff}$
in Eq.~(\ref{eq:veff0}) in the limit that $2| M | \pi R > 1$. We have
\begin{equation}
  V_{eff} \simeq \frac{2 N_H}{2} \, \int\!\! \frac{d^4p}{(2 \pi)^4} \, 
   \text{ln} \left[ 1 + 
    \frac{ 4 \, \left( p^2 + M^{2} \right)}{ p^2 } 
  \sin^{2}( \omega \pi ) \, 
   e^{- 2 M \, \pi \, R \, \sqrt{ 1 + \frac{ p^2 }{ M^{2} } } }
    \right] \quad ;
     \label{eq:wexp}
\end{equation} 
which can be computed analytically by expanding the logarithm to give 
\begin{equation}
  V_{eff} \simeq \frac{ 2 N_H }{ 32 \pi^6 R^4 }
    \sin^{2}( \omega \pi ) \ F(M\pi R) \quad ;
 \label{eq:veff0exp}
\end{equation}
where
\begin{equation}
 F(x)= e^{- 2 x}\left[3+6x+6x^2+4x^3\right] \quad .
\label{eq:eFe}
\end{equation}
We have checked numerically that (\ref{eq:veff0exp}) is a good
approximation of (\ref{eq:veff0}) when $2|M|\pi R >1$.

\section{\sc Dynamical determination of SS-parameter}  \label{sec:SS}

The effective potential in (\ref{eq:veff0exp}) arising from bulk
fields with an odd parity mass has a minimum at $\omega=0$ and a
maximum at $\omega=1/2$.  However, when this is combined with the
potential (\ref{pot0}) generated by bulk fields {\it without} bulk
masses for the cases where it has a minimum at $\omega=1/2$ and a
maximum at $\omega=0$, the resulting total potential can have a global
minimum at intermediate values $0<\omega<1/2$. In fact if we have a
situation where the number of hypermultiplets with zero bulk masses
$N_h$ is such that $N_h<2+N_V$ and the potential (\ref{pot0}) has a
minimum at $\omega=1/2$, then by adding $N_H$ hypermultiplets with a
common mass $M$, the critical value of the mass $M^*$ for which
$\omega=1/2$ becomes a maximum and the minimum is shifted to
$\omega<1/2$ is provided by the solution of the following equation,
\begin{equation}
\label{ecuacion}
9(2+N_V-N_h)\,\zeta(3)=4 N_H\ F(M^*\pi R) \quad ;
\end{equation}
which is valid for values $|M|\pi R\simgt 1$ for which the
approximation leading to (\ref{eq:veff0exp}) holds. For instance,
consider the case where all three generations of quarks and leptons
live in the bulk with a common mass $M>0$.  From Eq.~(\ref{ecuacion})
we find that the minimum at $\omega=1/2$ is destabilized for values of
$M^*R\simeq 0.78$.
\begin{figure}[htb]
\begin{center}
\begin{picture}(300,200)(0,0)
\vspace{.5cm}
\psfrag{a}{(a)}
\psfrag{b}{(b)}
\psfrag{c}{(c)}
\psfrag{d}{(d)}
\psfrag{o1}{\bf $\omega $}
\psfrag{o2}{\bf $\omega $}
\psfrag{o3}{\bf $\omega $}
\psfrag{o4}{\bf $\omega $}
\psfrag{v1}{\hspace*{-8mm}$V_{eff}\,R^4$}
\psfrag{v3}{\hspace*{-8mm}$V_{eff}\,R^4$}
\psfrag{v2}{}
\psfrag{v4}{}
\epsfig{file=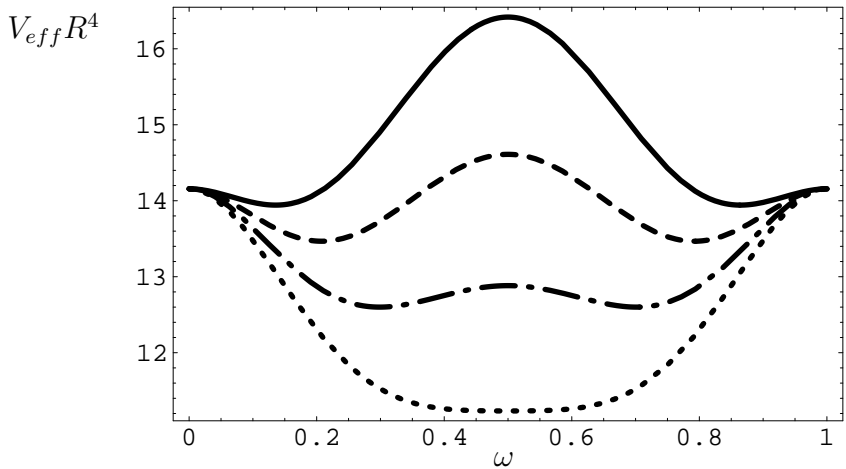}
\Text(-300,170)[c]{$V_{eff}R^4$}
\end{picture}
\end{center}
\caption{\small Effective potential (in units of $10^{-4}$) with
$N_V=12$, $N_h=0$ and $N_H=45$ for values of $MR=0.72$ (full), $0.74$
(dashed), $0.76$ (dot-dashed) and $0.78$ (dots).}
\label{fig2}
\end{figure}

\begin{figure}[htb]
\begin{center}
\scalebox{.5}{
{\mbox{
\psfrag{w}{\LARGE $\omega$}
\psfrag{MR}{\hspace{-2cm}\Large $MR$}
\psfrag{title}{}
\epsfig{file=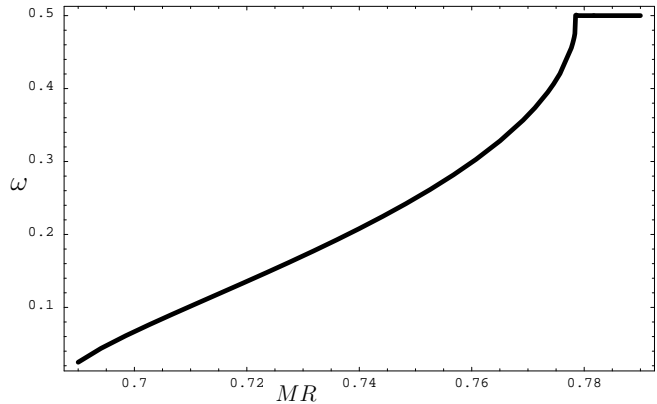}}}}
\caption{{\small Plot of $\omega$ as a function of $MR$ for the particle
content of Fig.~\ref{fig2}}}
\label{fig3}
\end{center}
\end{figure}

This is shown in Fig.~\ref{fig2} where the effective potential is
plotted as a function of $\omega$ for several values of $M$ and in
Fig.~\ref{fig3} where the minimum of $\omega$ is plotted as a function
of $MR$. Of course if there are several sets of hypermultiplets with
different masses (localizations) those with smaller masses (less
localized) provide the leading contribution to the effective
potential. For instance in the example above, localized states with
masses $MR\gg 1$ would not alter the dynamical minimization with
respect to $\omega$.

%
\section{\sc Conclusions}  \label{sec:conc}

In this paper we have shown that the SS-mechanism for supersymmetry
breaking in five-dimensional orbifold theories is affected by the
level of localization of five-dimensional fields on
boundaries. Indeed, the SS supersymmetry breaking parameter $\omega$
turns out to be a function of the localizing mass term $M$. The value
of the VEV of the parameter $\omega$ is fixed by one-loop corrections
and, in the absence of quasi-localized five-dimensional fields, is
fixed to be either 0 or 1/2 depending on the number of bulk
hypermultiplets and vector multiplets. However, with quasi-localized
fields, the VEV of the SS parameter can assume any intermediate
value. 

Our results can be generalized to the case in which the
five-dimensional symmetry is a gauge symmetry. This is particularly
interesting in theories where the Standard Model Higgs boson can be
identified with the extra dimensional component of a gauge boson and
the higher dimensional gauge group is broken to the Standard Model
group by the orbifold action.  In that case the Standard Model
symmetry can be radiatively broken by the Hosotani mechanism and the
Higgs boson mass is protected from bulk quadratic divergences by the
higher dimensional gauge theory without any need for
supersymmetry~\cite{higgsextra}.  We are presently investigating how
our results can be extended to such theories in order to reproduce
satisfactory Yukawa couplings and Higgs potentials.

\vspace*{7mm}
\noindent {\bf \Large Acknowledgements}

\noindent This work was supported in part by the RTN European Programs
HPRN-CT-2000-00148 and HPRN-CT-2000-00152 and by CICYT, Spain, under
contracts FPA 2001-1806 and FPA 2002-00748. The work of GG was
supported by DAAD. AR thanks IFAE for hospitality during the
completion of this work.



\begin{thebibliography}{99}
\bibitem{hosotani}
Y.~Hosotani,
Phys.\ Lett.\ B {\bf 126} (1983) 309;
%
%
{\it ibidem}, 
Phys.\ Lett.\ B {\bf 129} (1983) 193; 
%
%
{\it ibidem},
Annals Phys.\  {\bf 190} (1989) 233.
\bibitem{ss} 
J.~Scherk and J.~H.~Schwarz,
Phys.\ Lett.\ B {\bf 82} (1979) 60;
%
%
{\it ibidem}, 
Nucl. Phys. {\bf B153} (1979) 61. 
\bibitem{explicit} 
A.~Pomarol and M.~Quiros,
Phys.\ Lett.\ B {\bf 438} (1998) 255
[arXiv:hep-ph/9806263];
%
I.~Antoniadis, S.~Dimopoulos, A.~Pomarol and M.~Quiros,
Nucl.\ Phys.\ B {\bf 544} (1999) 503
[arXiv:hep-ph/9810410];
%
R.~Barbieri, L.~J.~Hall and Y.~Nomura,
Phys.\ Rev.\ D {\bf 63} (2001) 105007
[arXiv:hep-ph/0011311];
%
N.~Arkani-Hamed, L.~J.~Hall, Y.~Nomura, D.~R.~Smith and N.~Weiner,
Nucl.\ Phys.\ B {\bf 605} (2001) 81
[arXiv:hep-ph/0102090];
%
A.~Delgado and M.~Quiros,
Nucl.\ Phys.\ B {\bf 607} (2001) 99
[arXiv:hep-ph/0103058];
%
A.~Delgado, G.~von Gersdorff, P.~John and M.~Quiros,
Phys.\ Lett.\ B {\bf 517} (2001) 445
[arXiv:hep-ph/0104112];
%
R.~Contino and L.~Pilo,
Phys.\ Lett.\ B {\bf 523} (2001) 347
[arXiv:hep-ph/0104130];
%
Y.~Nomura,
arXiv:hep-ph/0105113;
%
N.~Weiner,
arXiv:hep-ph/0106021;
%
%
R.~Barbieri, L.~J.~Hall and Y.~Nomura,
Nucl.\ Phys.\ B {\bf 624} (2002) 63
[arXiv:hep-th/0107004];
%
%
A.~Masiero, C.~A.~Scrucca, M.~Serone and L.~Silvestrini,
Phys.\ Rev.\ Lett.\  {\bf 87} (2001) 251601
[arXiv:hep-ph/0107201];
%
%
A.~Delgado, G.~von Gersdorff and M.~Quiros,
Nucl.\ Phys.\ B {\bf 613} (2001) 49
[arXiv:hep-ph/0107233];
%
%
M.~Quiros,
J.\ Phys.\ G {\bf 27} (2001) 2497;
%
%
V.~Di Clemente, S.~F.~King and D.~A.~Rayner,
Nucl.\ Phys.\ B {\bf 617} (2001) 71
[arXiv:hep-ph/0107290];
%
%
V.~Di Clemente and Y.~A.~Kubyshin,
Nucl.\ Phys.\ B {\bf 636} (2002) 115
[arXiv:hep-th/0108117];
%
%
D.~M.~Ghilencea, H.~P.~Nilles and S.~Stieberger,
New J.\ Phys.\  {\bf 4} (2002) 15
[arXiv:hep-th/0108183];
%
%
H.~D.~Kim,
Phys.\ Rev.\ D {\bf 65} (2002) 105021
[arXiv:hep-th/0109101];
%
T.~Gherghetta and A.~Riotto,
Nucl.\ Phys.\ B {\bf 623} (2002) 97
[arXiv:hep-th/0110022];
%
V.~Di Clemente, S.~F.~King and D.~A.~Rayner,
Nucl.\ Phys.\ B {\bf 646} (2002) 24
[arXiv:hep-ph/0205010].
\bibitem{geromariano}
G.~von Gersdorff and M.~Quiros,
Phys.\ Rev.\ D {\bf 65} (2002) 064016
[arXiv:hep-th/0110132].
\bibitem{Quiros:2003gg} For a review, see: M.~Quiros, ``New ideas in
symmetry breaking,'' Lectures given at TASI 2002, Boulder, Colorado,
2-28 Jun 2002, arXiv:hep-ph/0302189.
\bibitem{anomalies}
G.~von Gersdorff and M.~Quiros,
arXiv:hep-th/0305024.
\bibitem{geromarianotoni}
G.~von Gersdorff, M.~Quiros and A.~Riotto,
Nucl.\ Phys.\ B {\bf 634} (2002) 90
[arXiv:hep-th/0204041].
\bibitem{georgi} 
H.~Georgi, A.~K.~Grant and G.~Hailu,
Phys.\ Rev.\ D {\bf 63} (2001) 064027
[arXiv:hep-ph/0007350];
%
%
N.~Arkani-Hamed, A.~G.~Cohen and H.~Georgi,
Phys.\ Lett.\ B {\bf 516} (2001) 395
[arXiv:hep-th/0103135].
%
%
\bibitem{nilles}
D.~M.~Ghilencea, S.~Groot Nibbelink and H.~P.~Nilles,
Nucl.\ Phys.\ B {\bf 619} (2001) 385
[arXiv:hep-th/0108184].
\bibitem{bar}
R.~Barbieri, R.~Contino, R.~Creminelli, R.~Rattazzi and C.~A.~Scrucca 
Phys.\ Rev.\ D. {\bf 66} (2002) 024025
[arXiv:hep-th/0203039];
%
%
D.~Marti and A.~Pomarol,
Phys.\ Rev.\ D {\bf 66} (2002) 125005
[arXiv:hep-ph/0205034].
\bibitem{appear}
G.~von Gersdorff, L.~Pilo, M.~Quir\'os, D.~A.~Rayner and A.~Riotto, 
in preparation.
\bibitem{quiros}
A.~Delgado, A.~Pomarol and M.~Quiros,
Phys.\ Rev.\ D {\bf 60} (1999) 095008
[arXiv:hep-ph/9812489].
\bibitem{rothstein}
W.~D.~Goldberger and I.~Z.~Rothstein,
hep-th/0208060.
\bibitem{higgsextra} 
I.~Antoniadis, K.~Benakli and M.~Quiros,
New J.\ Phys.\  {\bf 3} (2001) 20
[arXiv:hep-th/0108005];
%
%
G.~von Gersdorff, N.~Irges and M.~Quiros,
Nucl.\ Phys.\ B {\bf 635} (2002) 127
[arXiv:hep-th/0204223];
%
%
G.~von Gersdorff, N.~Irges and M.~Quiros,
arXiv:hep-ph/0206029;
%
%
C.~Csaki, C.~Grojean and H.~Murayama,
Phys.\ Rev.\ D {\bf 67} (2003) 085012
[arXiv:hep-ph/0210133];
%
%
G.~von Gersdorff, N.~Irges and M.~Quiros,
Phys.\ Lett.\ B {\bf 551} (2003) 351
[arXiv:hep-ph/0210134];
%
G.~Burdman and Y.~Nomura,
Nucl.\ Phys.\ B {\bf 656} (2003) 3
[arXiv:hep-ph/0210257];
%
C.~A.~Scrucca, M.~Serone and L.~Silvestrini,
arXiv:hep-ph/0304220.

\end{thebibliography}
\end{document}